\documentclass{bmcart}

\usepackage{amsthm,amsmath}
\RequirePackage{natbib}
\usepackage[utf8]{inputenc} 

\usepackage{graphicx}

\startlocaldefs
\endlocaldefs

\begin{document}

\begin{frontmatter}

\begin{fmbox}
\dochead{Research}


\title{CubeSat Quantum Communications Mission}


\author[
   addressref={aff1,aff13},                   
   corref={aff1},                       
   email={daniel.oi@strath.ac.uk}   
]{\inits{DKLO}\fnm{Daniel K L} \snm{Oi}}
\author[
   addressref={aff2,aff9},
   email={cqtalej@nus.edu.sg}
]{\inits{AL}\fnm{Alex} \snm{Ling}}
\author[
   addressref={aff3},
   email={vallone@dei.unipd.it}
]{\inits{GV}\fnm{Giuseppe} \snm{Vallone}}
\author[
   addressref={aff3},
   email={paolo.villoresi@dei.unipd.it}
]{\inits{PV}\fnm{Paolo} \snm{Villoresi}}
\author[
   addressref={aff10},
   email={steve.greenland@craftprospect.com}
]{\inits{SG}\fnm{Steve} \snm{Greenland}}
\author[
   addressref={aff10},
   email={emma.kerr.100@strath.ac.uk}
]{\inits{EK}\fnm{Emma} \snm{Kerr}}
\author[
   addressref={aff5},
   email={malcolm.macdonald.102@strath.ac.uk}
]{\inits{MM}\fnm{Malcolm} \snm{Macdonald}}
\author[
   addressref={aff6},
   email={harald.weinfurter@physik.uni-muenchen.de}
]{\inits{HW}\fnm{Harald} \snm{Weinfurter}}
\author[
   addressref={aff7},
   email={J.M.Kuiper@tudelft.nl}
]{\inits{JMK}\fnm{Hans} \snm{Kuiper}}
\author[
   addressref={aff11,aff12},
   email={e.charbon@tudelft.nl}
]{\inits{EC}\fnm{Edoardo} \snm{Charbon}}
\author[
   addressref={aff8},
   email={Rupert.Ursin@oeaw.ac.at}
]{\inits{RU}\fnm{Rupert} \snm{Ursin}}


\address[id=aff1]{
  \orgname{SUPA Department of Physics, University of Strathclyde}, 
  \street{John Anderson Building, 107 Rottenrow East},                     %
  \postcode{G4 0NG}                                
  \city{Glasgow},                              
  \cny{UK}                                    
}
\address[id=aff13]{%
	\orgname{Strathclyde Space Institute, James Weir Building, University of Strathclyde},
	\street{75 Montrose Street},
	\postcode{G1 1XJ}
	\city{Glasgow},
	\cny{United Kingdom}  
}
\address[id=aff2]{%
  \orgname{Centre for Quantum Technologies, National University of Singapore},
  \street{Block S15, Science Drive 2},
  \postcode{117543}
  \cny{Singapore}
}
\address[id=aff9]{%
  \orgname{Dept. of Physics, National University of Singapore},
  \street{2 Science Drive 3},
  \postcode{117542}
  \city{},
  \cny{Singapore}  
}
\address[id=aff3]{%
  \orgname{Dipartimento di Ingegneria dell'Informazione, Universit\`a degli Studi di Padova},
  \street{Via Giovanni Gradenigo, 6},
  \postcode{35131}
  \city{Padova},
  \cny{Italy}
}

\address[id=aff10]{%
  \orgname{Advanced Space Concepts Laboratory, Mechanical and Aerospace Engineering, University of Strathclyde},
  \street{James Weir Building, 75 Montrose Street},
  \postcode{G1 1XJ}
  \city{Glasgow},
  \cny{United Kingdom}  
}

\address[id=aff5]{%
  \orgname{Scottish Centre of Excellence in Satellite Applications},
  \street{Technology and Innovation Centre, 99 George Street},
  \postcode{G1 1RD}
  \city{Glasgow},
  \cny{United Kingdom}
}
\address[id=aff6]{%
  \orgname{Department f{\"u}r Physik, Ludwig-Maximilians-Universit{\"a}t},
  \street{Schellingstr. 4/III},
  \postcode{D-80799}
  \city{Munich},
  \cny{Germany}
}
\address[id=aff7]{%
  \orgname{Space Systems Engineering, Aerospace Engineering, Delft University of Technology},
  \street{Kluyverweg 1},
  \postcode{2629}
  \city{Delft},
  \cny{The Netherlands}
}

\address[id=aff11]{%
  \orgname{AQUA, EPFL},
  \street{Rue de la Maladière 71b},
  \postcode{Case postale 526, CH-2002 Neuch\^{a}tel 2}
  \city{Lausanne},
  \cny{Switzerland}
}

\address[id=aff12]{%
  \orgname{Delft University of Technology},
  \street{Mekelweg 4},
  \postcode{2628}
  \city{Delft},
  \cny{The Netherlands}
}

\address[id=aff8]{%
  \orgname{Institute for Quantum Optics and Quantum Information - Vienna Austrian Academy of Sciences},
  \street{Boltzmanngasse 3},
  \postcode{1090}
  \city{Vienna},
  \cny{Austria}  
}


\begin{artnotes}
\end{artnotes}

\end{fmbox}


\begin{abstractbox}

\begin{abstract} 

Quantum communication is a prime space technology application and offers near-term possibilities for long-distance quantum key distribution (QKD) and experimental tests of quantum entanglement. However, there exists considerable developmental risks and subsequent costs and time required to raise the technological readiness level of terrestrial quantum technologies and to adapt them for space operations. The small-space revolution is a promising route by which synergistic advances in miniaturization of both satellite systems and quantum technologies can be combined to leap-frog conventional space systems development. Here, we outline a recent proposal to perform orbit-to-ground transmission of entanglement and QKD using a CubeSat platform deployed from the International Space Station (ISS). This ambitious mission exploits advances in nanosatellite attitude determination and control systems (ADCS), miniaturised target acquisition and tracking sensors, compact and robust sources of single and entangled photons, and high-speed classical communications systems, all to be incorporated within a $10kg$ $6litre$ mass-volume envelope. The CubeSat Quantum Communications Mission (CQuCoM) would be a pathfinder for advanced nanosatellite payloads and operations, and would establish the basis for a constellation of low-Earth orbit trusted-nodes for QKD service provision.

\end{abstract}


\begin{keyword}
\kwd{CubeSat}
\kwd{quantum}
\kwd{entanglement}
\kwd{cryptography}
\end{keyword}


\end{abstractbox}
%

\end{frontmatter}



\section{Introduction}

Quantum technologies are advancing at a rapid rate, with quantum key distribution (QKD) for secure communication being the most mature. Current fibre-based systems are best suited for short-range (a few $100km$) applications due to fibre attenuation restricting the maximum practical distance~\footnote{The development of quantum memories for quantum repeaters is a long-term solution to this short range but is far from maturity~\cite{boone2015entanglement}.}. Free-space optical transmission is another option but limited sight lines and horizontal atmospheric density again restricts its range. Satellite-based QKD systems have been proposed for establishing inter-continental QKD links~\cite{bacsardi2013way}. Feasibility of different aspects of the concept have been demonstrated by Earth-based experiments such as the transmission of quantum entanglement over $144km$~\cite{ursin2007entanglement}, performing QKD from an aircraft to ground~\cite{nauerth2013air}, ground to air~\cite{pugh2016airborne}, receiving single photons from retroreflectors in orbit~\cite{villoresi2008experimental,yin2013experimental,vallone2014experimental} and other moving platforms~\cite{wang2013direct,bourgoin2015free}. Various groups around the world are working towards space-based demonstrations of quantum communication~\cite{Morong2012,NanoQEY2014,scheidl2013quantum,scheidl2012space,elser2015satellite} but most have not been successfully launched. Only recently, the $600kg$ Quantum Experiments at Space Scale (QUESS) Satellite was launched on 17 August 2016, at 17:40 UTC by the China National Space Agency~\cite{wu2014strategi,quesslaunch}.

\begin{figure}
\includegraphics[width=0.9\textwidth]{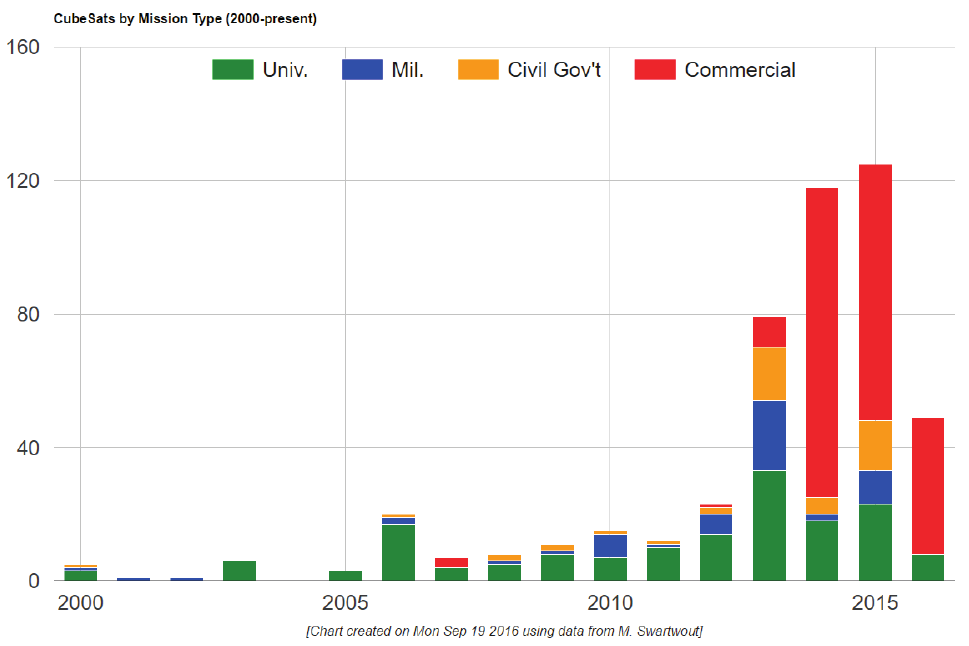}
\caption{CubeSat Launches. Since 2000, the rate of CubeSat launches has increased tremendously, especially in the last three years. The rate of university/educational CubeSat launches has been fairly steady, the recent growth has been driven by applications such as Earth observation and communications/tracking. Note: 2016 data incomplete.}
\label{fig:swartout}
\end{figure}

A barrier to experimental progress in this area has been the challenge of translating terrestrial quantum technology to the space environment, particularly in the context of the traditional ``big-space'' paradigm of satellite development and operations. This is characterized by large, long-term, high performance spacecraft with redundant systems following conservative design practice driven in part by the high cost of launch and satellite operations~\footnote{For example, Gravity Probe B cost USD750M and took over 50 years of development~\cite{reich2011troubled}, whilst the Hubble Space Telescope cost USD4.7B to launch~\cite{JWTICRP} and 20 years of development though these represent extreme examples of large space missions.}. A new paradigm has arisen, ``Micro-Space'' as embodied in the CubeSat standard~\cite{heidt2000cubesat}, that upturns the satellite development process. This approach exploits contemporary developments in miniaturization of electronics and other satellite systems to allow the construction and operation of highly capable spacecraft massing in the kilogram range, so-called nanosats~\footnote{The zoology of satellite classes includes mediumsats ($500-1000kg$) minisats ($100-500kg$), microsats ($10-100kg$), nanosats ($1-10kg$), picosats ($0.1-1kg$), and femtosats ($<0.1kg$) as well as large sats ($>1000kg$).}. In contrast, a geostationary communication satellite is typically 1000 times greater in mass. As cost of development, launch, and operations scales with mass, nanosatellites offer access to space at a vastly reduced cost that is affordable by small companies and research groups~\cite{shao2013performance}. The CubeSat standard was itself originally designed with undergraduate engineering educational projects in mind. Since the establishment of the CubeSat standard in 2000, it has become a very popular class of satellite ranging from hobbyists~\cite{wuerl2015lessons}, some countries first spacecraft~\cite{latt2014estcube}, basic space science~\cite{muylaert2009qb50}, to commercial services such as Earth imaging~\cite{foster2015orbit} and asset tracking~\cite{sarda2010canadian}. The standardized nature of the CubeSat platform has attracted commercial support for components and subsystems. It is possible to order online all the parts needed to assemble a fully functional CubeSat including structures, power systems, communications, ADCS, control, as well as basic payloads such as imagers. CubeSats are being launched in great numbers with over 120 launched in 2015, and 118 in 2014, with the proportion of commercial, scientific, and governmental use now the majority~\cite{Swartout2015} showing the transition from a purely educational tool to a valid applications platform (Fig.~\ref{fig:swartout}).

The role of CubeSats~\footnote{For the purposes of this article, we will use the terms CubeSat and nanosat interchangeably.} for space quantum technologies is two-fold~\cite{CPCubeSats2016}: firstly in the short term for pathfinder, technology demonstration, and derisking missions; secondly in the long-term for service provision for certain applications. CubeSats are not a panacea but their advantages of lower-cost, shorter development times, rapid and multiple deployment opportunities may be valuable for making more rapid progress in space quantum technologies. The CubeSat Quantum Communications Mission (CQuCoM) has been proposed to achieve at low cost and development time the key milestone of transmission of quantum signals from an orbiting source to a ground receiver. The goals are to perform quantum key distribution and the establishment of entanglement between space and ground. The mission would also represent a leap in capability for nanosatellites, especially for pointing and for carrying fundamental physics experiments. It is an extremely challenging project that stacks a number of critical systems engineering fields, in particular the combination of extreme high pointing accuracy and subsequent ADCS requirements and interactions. We present the mission concept, the key challenges, and outline the systems to be developed to overcome them.

\section{Mission CONOPS}

\begin{figure}
\includegraphics[width=0.9\textwidth]{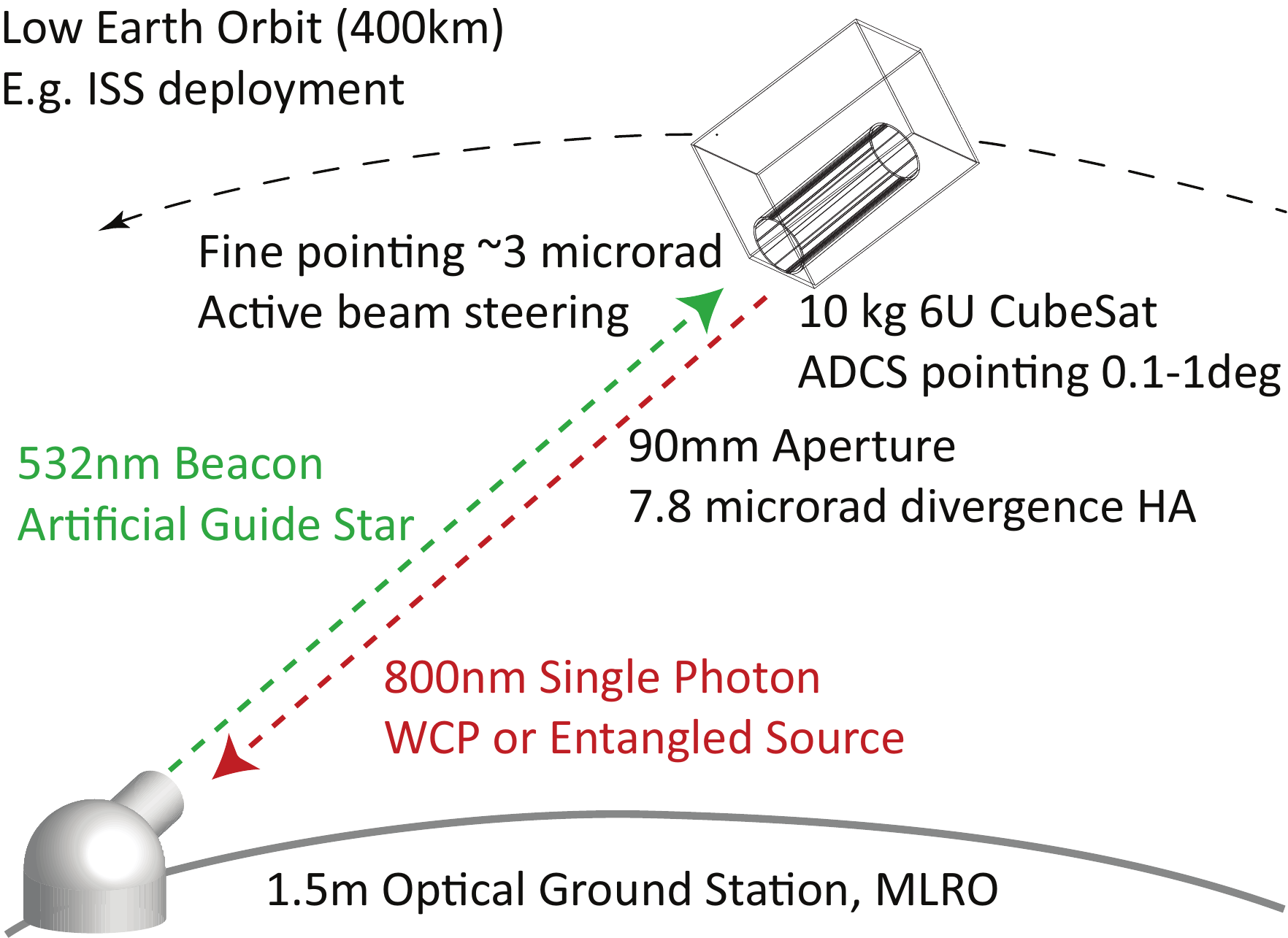}
\caption{Concept of Operations for CQuCoM. The CQuCoM 6U CubeSat would be deployed from the ISS into a circular low-Earth orbit. The ground track includes the Matera Laser Rangefinding Observatory operated by the University of Padua that would act as the optical ground station (OGS). The OGS would transmit a strong guide beacon at 532nm allowing the CQuCoM CubeSat to acquire and begin tracking the target position. Rotating the entire satellite to point towards the OGS provides coarse pointing to sub-degree level, sufficient to bring the OGS beacon within the acquisition field of view of the beacon tracking sensor. The beacon tracker is co-aligned with the outgoing signal photons and allows precision determination of the direction of transmit telescope boresight. The error signal from the beacon tracker is used to drive fast-steering mirrors to direct signal photons to the OGS. The fine-pointing system takes into account the velocity aberration with point-ahead correction. A quantum source on board the satellite provides single-photon level signals that are detected by the OGS. A switchable strong/weak coherent pulse source allows both the possibility of characterization of pointing performance and the free-space channel as well as quantum key distribution. An entangled-photon source would allow the distribution of entanglement between space and ground, one of the photon-pair is measured onboard and the results are compared with its respective partner detected on the ground.}
\label{fig:conops}
\end{figure}

The concept of operations (CONOPS) is presented in Fig.~\ref{fig:conops}. The basic task is to send quantum signals at the single photon level from an orbiting platform to a ground receiver. This paradigm was selected as it typically results in approximately 10 dB improvement in link loss when compared to the ground-to-space scenario. Two quantum sources are envisaged, a weak coherent pulse (WCP) source for performing a BB84-type QKD protocol, and an entangled photon pair source that would send one-half of each entangled photon pair to the ground receiver and retain for analysis the other half. The low-Earth orbit (LEO) reduces the losses due to range and simplifies space deployment, but introduces other challenges such as residual atmospheric disturbance. The major hurdle to overcome is the extremely high pointing accuracy required to minimize the link loss associated with free-space transmission over several hundred to a thousand kilometres.

The preliminary mission design calls for the launch of a 6U CubeSat~\footnote{A 1-unit (1U) CubeSat is nominally a $10cm$ cube of mass $1kg$. Several units can be combined to create CubeSats of greater mass, volume, and capability. Extensions to the standard allow for higher densities, up to $2kg$ per U~\cite{Hevner2011}.} to the International Space Station (ISS). An advantage of CubeSats (shared by other smallsats) is that it is delivered to the launch provider in a standardized container (deployer) format, such as PPOD or IPOD, that greatly simplifies the process of integration of the smallsat with the launch vehicle~\cite{swartwout2014first}. Regular resupply launches to the ISS gives greater mission flexibility for satellite development and operation. Commercial launch brokers provide streamlined access to space, a 6U CubeSat can be launched within 6 months of contract signing and for USD545K~\cite{spaceflight}. Baselining the ISS as a deployment platform removes uncertainty about orbital parameters and eases mission planning.

\section{CubeSat Platform}

The 6U platform was selected as it is the largest commonly handled CubeSat size whose cost/capability trade-off is favourable for many high-performance nanosatellite missions~\cite{TsitasKingston2012}. Several design studies have used 6U CubeSats for Earth observation as it can accommodate a reasonably large optical assembly together with ancillary payloads~\cite{turner2010nps,agasid2010study,straub2012smallsat}. Flown 6U missions include Perseus-M 1 \& 2 (19th June 2014 DNEPR), VELOX-II (6th December 2015 PSLV) and $^3$CAT-2 (15th August 2016) demonstrating system qualification compliance. There are approximately 65 6U missions under development. The use of CubeSats is not restricted to Earth orbit. A pair of 6U satellites, Mars Cube One, are to be used as interplanetary relay stations for the Mars lander InSight originally due for launch in 2016 (now scheduled for 2018 due to problems unrelated to the CubeSats)~\cite{skrobot2016cubesat}, demonstrating the capability that can be packed into this format.

An advantage of the CubeSat approach is the availability of conventional off-the-shelf (COTS) components in order to reduce costs and development time. The CQuCoM CubeSat will be based upon the PICosatellite for Atmospheric and Space Science Observations (PICASSO) platform developed by Clyde Space Ltd~\cite{mero2015picasso}. Though PICASSO is a 3U CubeSat, its systems can be used in a 6U structure with little modification. The platform provides an electrical power system (EPS), communications (COMMS), attitude determination and control systems (ADCS), and an on-board computer (OBC). Integration of the payload with the platform would be performed using the NANOBED facility at the University of Strathclyde. We outline the key specifications of the CQuCoM platform below.

\subsection{Structure}
\begin{figure}
\includegraphics[width=0.9\textwidth]{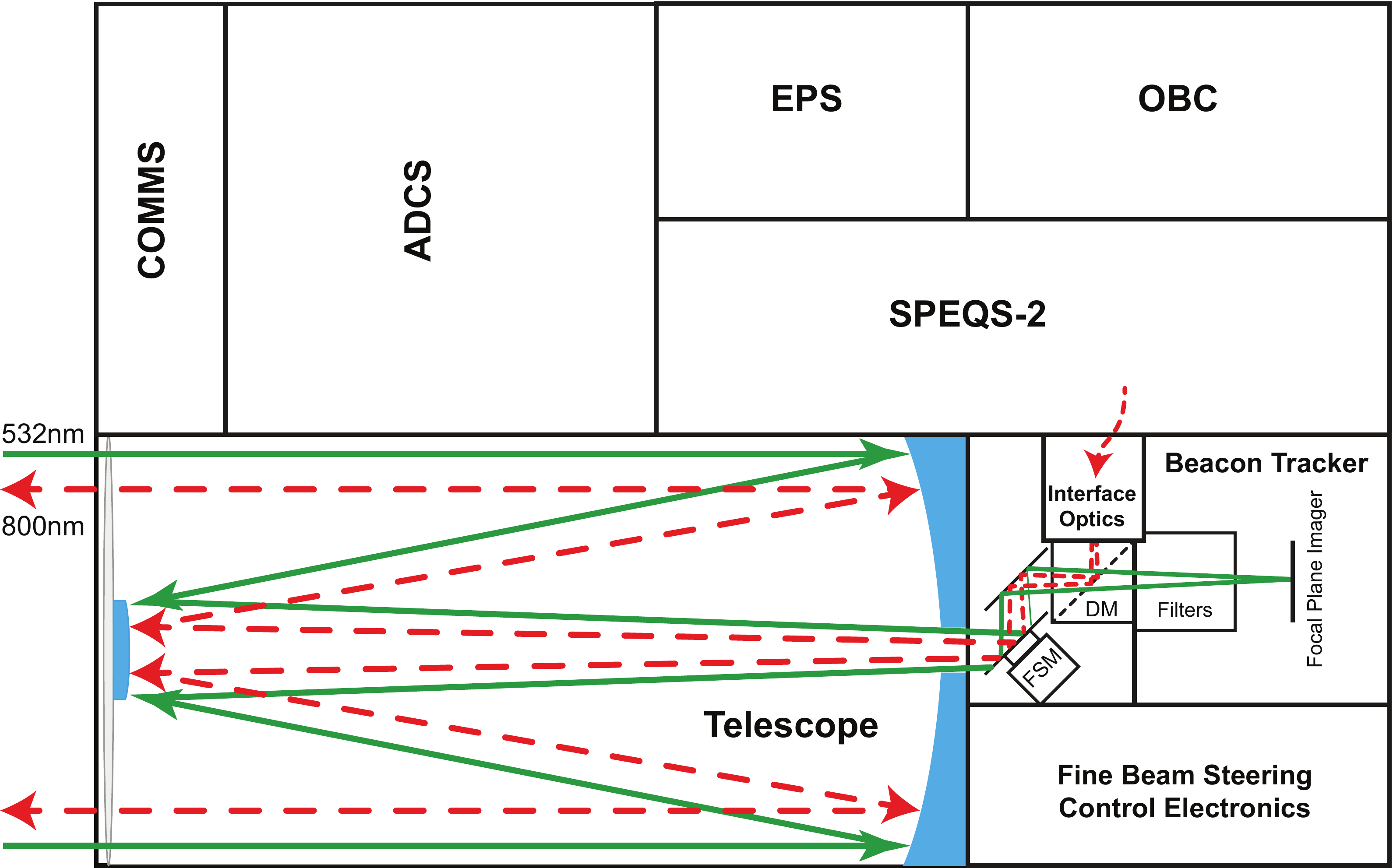}
\caption{CQuCoM CubeSat Layout. One half of the structure is devoted to the transmission optics which includes a telescope, beacon tracker, beam steering, and optical interface with the quantum source. The platform systems (COMMS, ADCS, EPS, and OBC) are based upon the PICASSO 3U CubeSat developed by Clyde Space and VTT Finland for the Belgian Institute of Space Aeronomy and ESA. Body mounted solar panels would provide power to the EPS for storage and distribution. Communications would be handled by UHF, S-band, and X-band radio systems. The ADCS consists of Earth, sun and star trackers, magnetorquers, and 3-axis momentum wheels. The OBC handles systems operations. Processing of data is performed in the mission computer as part of the entangled photon source (SPEQS-2) payload.}
\label{fig:schematic}
\end{figure}

These systems would be placed into a 6U (nominal $12cm\times 24cm\times 36cm$) structure~\cite{Hevner2011}. The CubeSat volumetric breakdown consists of 2U allocated to platform systems mentioned above, 1U to the quantum source, and 3U for the transmission optics (Fig.~\ref{fig:schematic}). Suitable 6U structures are available from a variety of vendors such as Innovative Solutions in Space~\cite{ISIS6U} and Pumpkin~\cite{Pumpkin6U}.

\subsection{EPS}
The satellite is powered by body-mounted solar panels that feed the EPS for storage and distribution of power. The low duty cycle of the transmission experiment eliminates the need for a deployable solar array reducing cost and complexity whilst increasing reliability. The lack of extraneous projected area also reduces the possibility of atmospheric buffeting. Orbit averaged power is 11W assuming 80\% sun-tracking efficiency. As transmission experiments are performed during eclipse, the EPS must be able to support the payload power draw using battery reserves alone. A $30WHr$ lithium-ion battery pack has been sized to support mission operations with sufficient depth of discharge margin to prevent cell degradation from repeated experimental runs. 
 
\subsection{COMMS}
Several radio systems are employed for (classical) communications. A UHF dipole array is used for tracking, telemetry, and control (TT{\&}C) and provides redundancy for low-speed data transmission ($100kb/s$). An S-band patch antenna is used for high-speed uplink (nominally $1Mb/s$). For high speed downlink of mission data, X-band CubeSat transmitters are commercially available and provide up to $100Mb/s$ data rate~\cite{syrlinks}. A GPS patch antenna is also incorporated into a face of the CubeSat. Space-rated GPS systems enable tracking of position and velocity to metre and sub-$m.s^{-1}$ accuracy respectively~\cite{kahr2011gps}. Onboard GPS enables precise orbital determination and calibration of two-line element measurements, necessary for the OGS to initially acquire the satellite and also for the ADCS to point the transmitter telescope towards the OGS to enable the optical beacon tracker (OBT) to lock onto the beacon sent up by the OGS.

\subsection{OBC}
 
The on-board computer is responsible for routine operations of the spacecraft. Low power space qualified processors and memory are available for CubeSats from a variety of vendors, typically based upon ARM devices and flash storage. The OBC will support different mission modes including initial switch-on and detumbling, charging, RAM attitude keeping, experiment, and data download modes. Failsafe modes including in-orbit reset will be included. A facility to update operational software is desirable as this allows experiments to be performed that were not envisaged prior to launch.
 
\subsection{ADCS}

The ADCS is used to provide coarse pointing by rotating the CubeSat body to align the transmitting telescope with the optical ground station during quantum transmission. The required level of ADCS accuracy has previously been challenging to achieve in nanosatellites due to a lack of high performance star trackers suitable for CubeSat applications. Only recently has there been commercial availability of such systems such as Blue Canyon Technologies XACT ACDS~\cite{mason2016minxss} with similar systems available from Maryland Aerospace~\cite{MAI} and Berlin Space Technologies~\cite{BST}. In particular, the aforementioned BCT XACT ADCS system has been demonstrated in-orbit pointing performance of $8$ arcseconds (1\textendash$\sigma$) on the MinXSS 3U CubeSat, this was independently verified by scientific instruments onboard. This level of pointing accuracy indicates that CubeSats can now seriously be considered for missions requiring precision pointing.

The PICASSO ADCS system upon which the CQuCoM satellite is based provides $<1^\circ$ pointing accuracy. A full system engineering analysis will determine whether this baseline level of pointing is sufficient for the CQuCoM mission, the BCT XACT platform is a viable alternative should higher accuracy coarse pointing be required. The ADCS utilizes a combination of sensors such as a 3-axis magnetometer to detect the strength and orientation of the Earth's magnetic field, and angular rate sensors to measure the rotational velocity of the satellite. To establish absolute attitude, coarse and fine Sun sensors are used when sunlit but during eclipse, when experimental transmission occurs these sensors are ineffective. Instead, a high precision star tracker is used to provide accurate 3-axis pointing knowledge. Attitude control is through a combination of magnetic torque actuators (magnetorquers or MTQs) interacting with the Earth's magnetic field, and reaction wheels. The MTQs are used for detumbling and for desaturating the reaction wheels. 

\section{Quantum Sources and Detectors}

The CQuCoM proposal involves two missions with different quantum sources. The first mission will validate the transmission system. Numerical studies of the optical channel between space and ground predict a link loss of $-30$ or $-40 dB$ for a spacecraft with a $10cm$ aperture at $500 km$ altitude and a $1m$ aperture at the optical ground station~\cite{bourgoin2015free}. As CQuCoM will be at a lower altitude, it is imperative to establish first that the fine-pointing mechanism can overcome any residual atmospheric buffeting and greater traversal speed. The second mission would incorporate lessons learned from the first in performing the more challenging task of entanglement distribution.

Currently, the CQuCoM proposal calls for two sequential missions. It is possible, however, to consider the possibility of combining both missions into a single spacecraft. This will require the spacecraft to be able to supply more resources. For one, an increased volume for accommodating both types of light sources must be available. It also makes the optical interfaces more challenging.

\subsection{Weak Coherent Pulse Source}

To conduct the space-to-ground test, the first mission will use a modulated laser transmitter whose intensity can be tuned to act either as a strong optical beacon, or as a weak coherent pulse (WCP) source, where the average number of photons per pulse is much less than one. When acting as the strong optical beacon, it is possible to use this light source to characterize the space-to-ground optical channel and to commission the fine-pointing mechanism~\cite{nauerth2013airthesis}. When this is completed, the light source can be adjusted to become a polarisation-encoded WCP source that can carry out quantum key distribution using conventional prepare-and-send methods including decoy state protocols to prevent photon number splitting attacks.

WCP sources are well developed and have been miniaturized to fit within hand-held devices ($35\times 20\times 8mm^3$~\cite{melen2016integrated}) and represents a low-risk quantum signal source for the first mission. A true random number generator (RNG) would be required to guarantee security but the 1U set aside for the source should give ample payload margin~\footnote{High speed quantum RNGs have been demonstrated with suitable SWaP characteristics. For example, in~\cite{shi2016random} random generation at $480Mb/s$ was shown in a 0.1U package consuming a few watts, easily scalable to $846Mb/s$ or even higher. A chipscale QRNG component operating at $1Gb/s$ has been reported~\cite{abellan2016quantum}. For testing purposes, a pseudo-RNG could be used, or else random settings could be pre-computed.}. A baseline transmission rate of $100MHz$ with 0.5 photons/pulse should allow the generation of secure keys during a ground pass, with the option of increasing the rate to overcome additional link losses~\cite{bourgoin2015experimental}.

\subsection{Entangled Source SPEQS}

\begin{figure}
\includegraphics[width=0.8\textwidth]{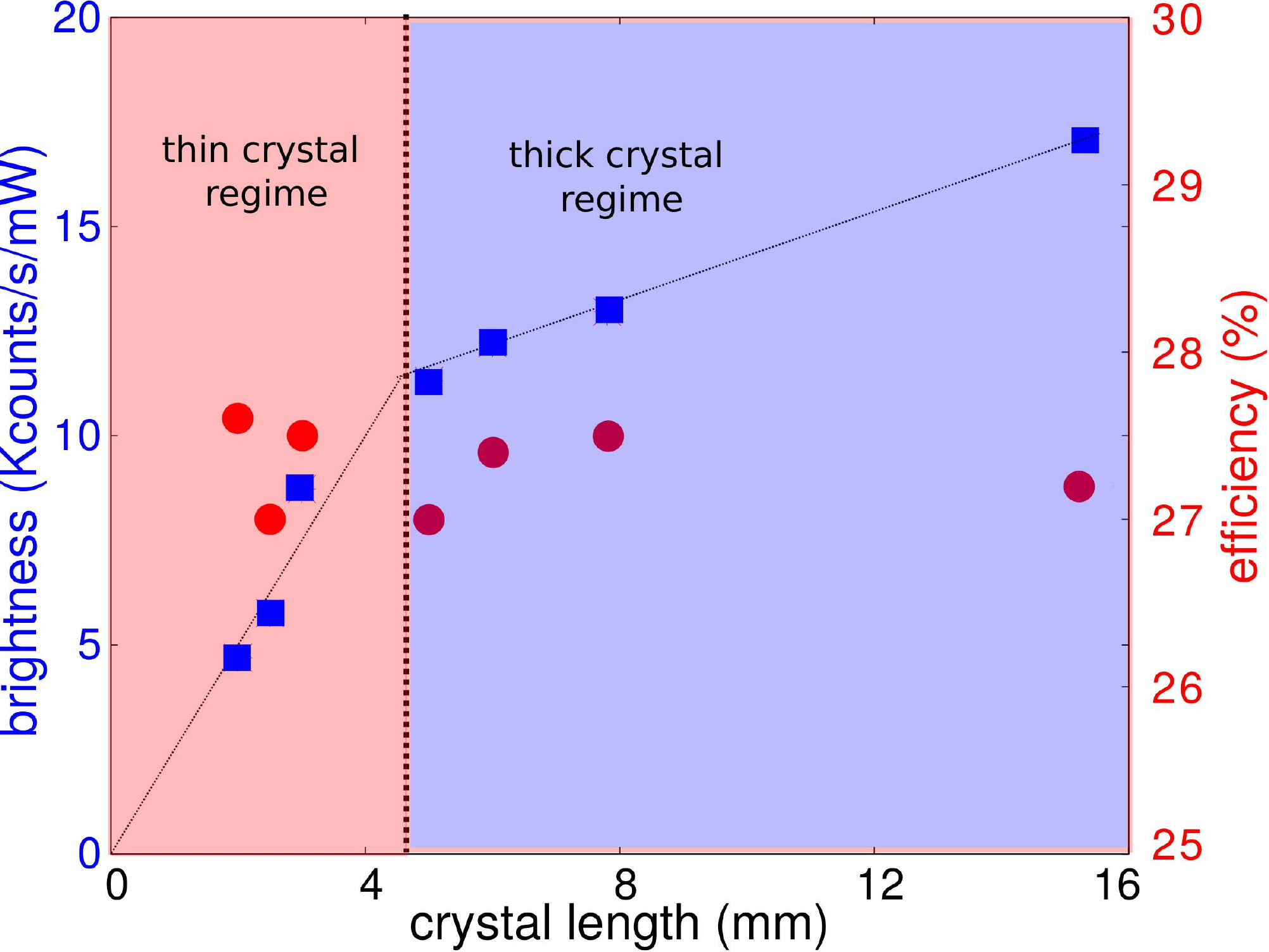}
\caption{The collection efficiency of SPDC photons (pair-to-singles ratio), and the final brightness in the SPEQS geometry, for different crystal lengths. In this graph, the pump and collection beam FWHM were fixed at 180 micron and 120 micron respectively. The collection efficiency is quite stable across the range of crystal lengths. The dependence of brightness on crystal length, however, appears to lie in two different regimes. Additional work is ongoing to characterize this dependence and a model is being developed.}
\label{fig:cl}
\end{figure}

The second CQuCoM mission will attempt entanglement-based QKD. The use of quantum entangled photon pairs has certain technical advantages over the more conventional prepare-and-send schemes. For example, a true random number generator is not required for the source as the measurement of entangled photons generate intrinsic randomness. Another interesting advantage is that the photon pairs, generated in a nonlinear optical process are created within femtoseconds of each other and it is possible to carry out time-stamping and correlation matching without the use of atomic clocks or GPS-type signals \cite{ho09}. Thus, entanglement-based systems in space have other interesting technology applications beside QKD.

The polarization-entangled source for CQuCoM is based on the Small Photon-Entangling Quantum System (SPEQS) currently designed and built at the National University of Singapore. The SPEQS devices, for the generation and detection of entangled photon pairs, are designed to be rugged and compact as it has to be contained within the size, weight and power (SWAP) constraints of nanosatellites~\cite{cheng15}. A notable feature of SPEQS devices is that they appear to be incredibly rugged, with one copy surviving the explosion of a space launch vehicle intact and in good working order \cite{tang15}. The first generation SPEQS devices have been space qualified, first through demonstration in near-space~\cite{tang14}, then formal testing after integration into nanosatellites, and finally through successful operation in orbit on the Galassia 3U CubeSat~\cite{tang2016generation,chandrasekara2016correlated}.

The polarization-entangled photon pairs are generated via spontaneous parametric downconversion (SPDC). The source geometry is based on collinear, Type-I, non-degenerate SPDC using bulk $\beta$-Barium Borate crystals for downconversion. The advantages of using BBO are that it is uniaxial and its optical properties (birefringence) are very temperature  tolerant. The single photons are currently detected by silicon Geiger-mode avalanche photodiodes (Si-APDs). Careful characterization studies show that the Type-I geometry enables a very robust set of pump and collection conditions that simultaneously achieve high pair rate (brightness) and a high pair-to-singles ratio. The length of the crystals is an important consideration. With fixed pump and collection beam parameters, the dependence of brightness on crystal length falls into two different regimes (see Fig.~\ref{fig:cl}). A trade-off in the target brightness and size of the source needs to be made~\cite{septriani2016thick}.

The entangled photon source that is being proposed for CQuCoM, called SPEQS-2, is currently being built at the NUS and is expected to consume about $10W$ of continuous power and to have a mass of about $500g$~\cite{chandrasekara15_spie2}. A separate qualification mission is being planned and the satellite mission and the SPEQS-2 detailed design specifications are described in an accompanying article~\cite{bedington2016nanosatellite}.

\subsection{Single Photon Detectors}

Due to the large downlink transmission losses, achieving a high enough entangled pair coincidence rate between the OGS and the CubeSat requires a high pair-production rate onboard CQuCoM, consequently we need high-speed single photon counters. Si-APDs are baselined for the second mission but we would also investigate the use of more advanced solutions to allow for faster pair generation that could not be easily handled by conventional Si-APDs due to timing resolution, jitter, deadtime, or power limitations.

Geiger-mode APDs or single-photon avalanche diodes (SPADs) can also be implemented in complementary metal-oxide semiconductor (CMOS) technologies, where the detectors are replicated in very large numbers on a single CMOS chip~\cite{veerappan2016low} and even in stacked CMOS chips~\cite{pavia20151}. The advantage is to be able to detect single-photons with very high single-photon time resolution in multiple locations, so as to minimize the dead time of the measurement. Another advantage of parallel detection is the capability of implementing multiple channels and thus incrementing the throughput of free-space quantum communications channels using space-division multiple access (SDMA) mechanisms. 

Thanks to Moore’s Law, it becomes possible to create complex digital signal processing on chip side-by-side with, or under the detectors, thus minimizing noise and jitter. Proximity of detection and processing maximizes compactness, while reducing power dissipation due to the lack of expensive and power-hungry drivers. This feature may be of significant value whenever power and space are in high demand, such as in satellites~\cite{charbon2014single}. CMOS SPADs have also shown resilience to gamma radiation and proton bombardment at several energies and doses, thus proving their suitability for space applications~\cite{maruyama20141024,charbon2010radiation}.

We have developed large linear arrays of SPADs with a diameter of several microns that exhibit a single photon timing resolution better than $100ps$ and a dead time, individually, of several tens of nanoseconds~\cite{burri2016linospad}. The arrays are coupled with digital hardware including time-to-digital converters (TDCs) capable of resolutions better than $25ps$ and recharge periods shorter than $7.5ns$. With these devices, it is possible to achieve overall deadtimes of several tens of picoseconds, while dissipating less than $100mW$. Thanks to parallelism of SPADs and TDCs, large throughputs of up to $34Gb/s$ can thus be achieved, while generally only several $Mb/s$ are exploited in single-photon communication.

\section{Ground Segment and Optical Ground Station}

Command and control of CQuCoM will be performed by a network of RF ground stations located in Glasgow (University of Strathclyde, mission control), Singapore (National University of Singapore), and Delft (TU Delft). The diversity of ground stations allows more frequent contact and greater opportunities for downlink of data. Mission control will also link with the OGS to co-ordinate experimental passes.

The CQuCoM satellite will transmit quantum signals (WCP or single photons) to an optical ground station located at the Matera Laser Ranging Observatory (MLRO), Italy. This facility has already conducted proof-of-principle quantum communication experiments utilizing laser signals bounced off retroreflectors mounted on existing satellites~\cite{Vallone2016PRL}. Essentially the same experimental setup will be used for the CQuCoM mission with the addition of an $532nm$ optical beacon~\footnote{MLRO has a two-colour laser rangefinding system at $532nm$ and $355nm$. This suggests utilising the existing $532nm$ laser systems for the beacon and the $355nm$ laser for rangefinding to avoid interference.}. A radio link at the OGS will be used to communicate with and monitor the satellite during the experiments. The ADCS telemetry will be used to align the measurement bases by polarization control at the OGS.

\subsection{Pass analysis}

\begin{figure}
\includegraphics[width=0.9\textwidth]{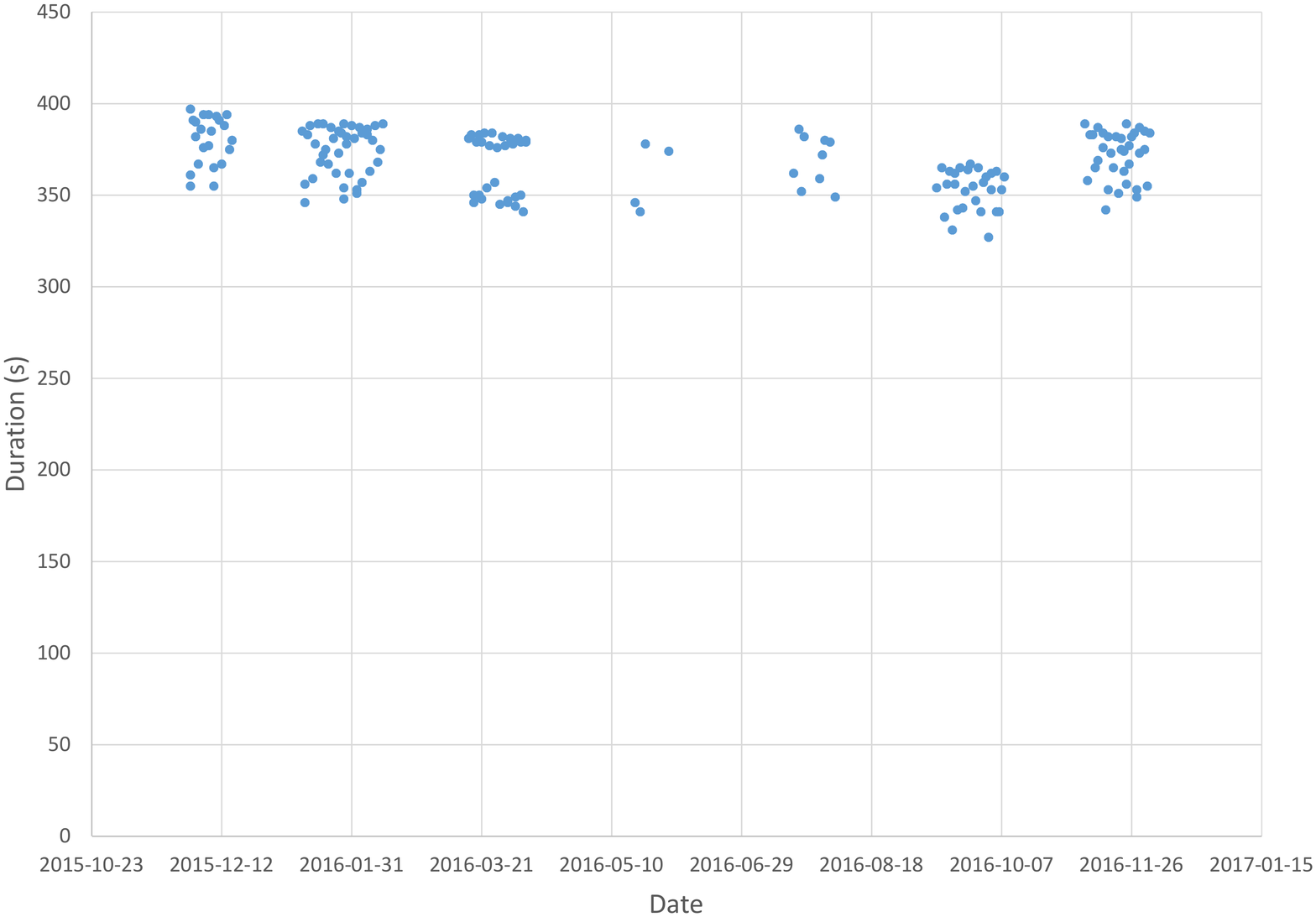}
\caption{12 Month Pass Analysis of the ISS over MLRO. The distribution of durations for suitable passes over the OGS is shown. Passes were restricted to orbits rising at least $30^\circ$ above the horizon and occurring in eclipse. The pass duration is counted as the time spent above $10^\circ$ above the horizon though the actual time available for quantum transmission will be less than this.}
\label{fig:passes}
\end{figure}

The baseline deployment from the ISS allows a preliminary determination of the orbital pass parameters for the selected OGS location of the MLRO. This is summarized in Fig.~\ref{fig:passes}, in a 12 month period, there are approximately 150 opportunities to conduct experimental operations between a satellite in the orbit of the ISS and OGS with an average pass duration of 6 minutes. We restrict transmission to night time when the satellite is also in eclipse to reduce background light entering the OGS receiver either from scattering of sunlight from the atmosphere or from reflected light off the satellite itself.

As the CubeSat has a lower ballistic co-efficient and does not carry any propellent to maintain altitude, the orbit will change and diverge from that of the ISS (which performs periodic orbit raising burns). At the initial deployment altitude of $400km$, a slant range of $1000km$ corresponds to minimum $23^\circ$ elevation, so we restrict ourselves to passes that rise to at least $30^\circ$ to allow sufficient time to perform initial acquisition and tracking. Passes that rise higher, and consequently for longer, will be used for transmission experiments. As the orbit of the CubeSat decays, pass opportunities and durations will reduce, though this will be partially compensated by the reduction in range leading to higher count rates at the OGS. We aim to perform experimental operations down to at least $300km$ altitude, below which atmospheric drag will quickly deorbit the spacecraft. A minimum experimental lifetime of 12 months should be achievable based on deorbit analysis in Section~\ref{sec:deorbit}.

\subsection{OGS Operations}

At the beginning of the transmission pass, the OGS would use orbital data, either two-line elements or GPS tracking data from onboard the CubeSat, to initialize the lock-on phase of the experiment. The OGS then sends rangefinding pulses that are sent back by retroreflectors mounted on the CubeSat allowing for both accurate distance determination (at the centimetre level) and tracking the precise direction of the CubeSat.

Once the OGS has found the CubeSat in its field of view, it can baffle the region of sky seen by the detector to reduce background stray light. The OGS will also transmit a laser beacon towards the CubeSat to guide its fine-pointing system.

The range information is used for time-of-flight timing correction between the transmitted pulses with measurements on the ground. For the WCP source, its pulsed nature allows windowing of the detection periods to reduce extraneous counts. This is not possible for the continuously pumped entangled photon source so coincidence matching will be used to precisely align the time-bases of the CubeSat with the OGS.

The $800nm$ wavelength of the quantum downlink allows the use of easily available Si-APDs for the OGS detectors. Moderate cooling is sufficient to reduce dark counts to negligible levels.

\section{Optics and Fine Pointing}

The main challenge of CQuCoM is the transmission of single photons from an orbital platform travelling at nearly $8kms^{-1}$ to the OGS. The CubeSat dimensions restrict the size of the transmission optics and the low mass constrains the pointing stability of the craft~\footnote{The MinXSS satellite has demonstrated $40\mu rad$ ($1\text{-}\sigma$) coarse pointing performance after being deployed from the ISS~\cite{mason2016minxss}.}. The transmission telescope diameter of $90mm$ will lead to a different beam divergence depending on the source~\cite{bourgoin2013comprehensive}. A WCP source allows a nearly flat wavefront to be transmitted leading to a divergence of $4\mu rad$ (HWHM) whilst the entangled photon pair source requires a $65mm$ Gaussian beam waist to optimize diffraction against truncation loss and this leads to a divergence of $7.8\mu$rad. The ground spot size varies from $3.2m$ (WCP source, Zenith) to $18m$ (entangled source, 20 degrees above horizon) for an orbital altitude of $400km$ leading to different geometric losses due to the finite collection aperture of the OGS. A fine-pointing specification of $3\mu rad$ has been chosen to balance the pointing losses against developmental cost and effort. The gains from a smaller pointing error diminish as the inherent divergence of the beam and other effects dominate.

The required pointing accuracy will be achieved by combining coarse (ADCS) and fine (OBT/BSM) pointing stages. The CubeSat will use 3-axis ADCS via reaction wheels for coarse pointing to aim its telescope at the OGS to within the acquisition FoV of the OBT to lock onto the OGS $532nm$ beacon laser. After initial lock, ADCS excursions up to the BSM FoV limit of several degrees can be accommodated.

\subsection{Transmission Optics}

The restricted size of a 6U CubeSat structure constrains the maximum optical aperture than can be easily employed. The use of deployable optics is being investigated by several groups~\cite{schwartz2016segmented, andersen2016falconsat,champagne2014cubesat,agasid2013collapsible} including at TU Delft with the Deployable Space Telescope project~\cite{dolkens2015deployable} together with TNO, ADS Leiden and ESA. However, a fixed optical system is attractive to minimize development risk. Planet employ $90mm$ Cassegrain-type reflector telescopes on their Dove 3U CubeSat constellation, 133 have been launched as of May 2016~\cite{colton2016supporting} thus demonstrating considerable flight heritage of this type of CubeSat optical system~\footnote{The optical system of the Planet ``flocks'' of ``doves'' has been refined over several generations: PS0 features a 2 element Maksutov Cassegrain optical system paired with an 11MP CCD detector. Optical elements are mounted relative to the structure of the spacecraft. PS1 features the same optical system as PS0, aligned and mounted in an isolated carbon fibre/titanium telescope. This telescope is matched with an 11MP CCD detector. PS2 features a five element optical system that provides a wider field of view and superior image quality. This optical system is paired with a 29MP CCD detector.~\cite{planet}}. 

As a baseline, we allocate 2U to the transmission telescope and its basic specifications are Cassegrain-type, $90mm$ diameter primary mirror, and $f=1400mm$ focal length. An athermal design can be used to minimize distortions due to temperature variations as the CubeSat moves into eclipse prior to any transmission experiment. The optical configuration will depend on the results of a trade-off study between manufacturing complexity/cost, optical performance, and compactness. Optical performance will depend on the ACDS coarse pointing accuracy that can be achieved as this drives the off-axis performance of the design to accommodate large BSM excursions. The combination of the optics, fine pointing and ADCS is an example of systems of systems engineering and this research would be an integral part of CQuCoM mission research and design.

\subsection{Beacon Tracker and Beam Steering}
 
Incoming 532nm beacon light sent from the OGS is separated from the outgoing beampath using a dichroic mirror, sent through an insertable narrow bandpass filter, to reduce stray light, and onward to the beacon tracker consisting of a modified star tracker~\footnote{A quadrant photodiode has typically been used in other beam steering experiments, or alternatively a 2-D tetra-lateral Position Sensitive Device (PSD). A modified startracker approach was chosen to allow for lock-on capture over a large field-of-view to mitigate against ADCS coarse pointing performance shortfalls. This also gives the possibility of obtaining imagery from the CubeSat for independent testing of pointing performance.}. During a frame, the defocussed image of the beacon is imaged onto a pixel array. The integration time is chosen to be short enough so that the image is not smeared. The deliberately defocussed point is spread across several pixels and the Gaussian intensity profile is determined from measurements of neighbouring pixels, a centroiding algorithm is then used to estimate the centre position of the beacon to sub-pixel accuracy. The accuracy by which this can be performed depends on the image signal to noise ratio (SNR) but better than $\frac{1}{40}$-pixel precision is achievable for moderate levels of noise and $\frac{1}{20}$-pixel for high levels of noise~\cite{delabie2013accurate}.
We will drive the OBT at a high frame rate ($\sim 300Hz$ full array readout, $\sim kHz$ with region-of-interest readout) in order to reduces the beacon frame interval and the possibility of image smear. To achieve sufficient SNR, the beacon power can be increased.

An attitude model for the satellite, with input from the OBT and high bandwidth inertial measurement units (IMUs), drives a beam steering mirror (BSM) for fine-pointing. Depending on the pass geometry and position of the satellite, the outgoing $800nm$ beam needs to be sent in a slightly different direction to the apparent position of the beacon due to velocity aberration~\footnote{The Doppler shift does not pose a problem for this mission. At ISS orbital speed of $7.67km s^{-1}$ the maximum wavelength shift is $0.02nm$, much smaller than the $0.1nm$ bandpass of ultra-narrow interference filters used for straylight rejection.}. The magnitude of the point-ahead correction can reach up to $54\mu$rad when passing over zenith. The ADCS is also sent the OBT/BSM offset so that the coarse pointing error can be closed, bringing the telescope boresight towards the beacon direction and reducing the possibility of the BSM exceeding its excursion limits.
 
\subsection{Pointing Errors}

Considering the interaction of the various sub-systems in determining pointing performance is a significant systems engineering challenge spanning all parties and disciplines. There are several potential sources of pointing error either leading to low frequency biases or high frequency noise in the transmitted beam direction.  Low frequency drift will misalign the telescope bore axis from the OGS direction and if left unchecked could bring the deviation outside of the angular limits of the beam steering mechanism (BSM). As long as the coarse pointing system can keep the optical boresight to within these limits, the final pointing performance will be determined mainly by the fine pointing mechanism. This will be mainly impacted by high frequency noise leading to jitter or beam wandering. A high optical OBT detection bandwidth is essential for rejecting this source of noise. Noise with higher frequency components than the OBT frame rate can be tackled by the IMUs and blended rate sensor fusion to compensate for any motion occurring in-between frames of the OBT~\cite{ortiz2001sub,gutierrez2011line}. Quantum communication experiments have achieved a few $\mu rad$ accuracy under demanding conditions such as in a propeller driven airplane~\cite{nauerth2013air} or lofted on a hot air balloon~\cite{wang2013direct}. The more benign microgravity environment and lower vibrational background of a space-based experiment should allow at least as good performance and we consider residual effects that may affect pointing performance.

\subsubsection{Solar pressure, Residual Magnetic Moment, and Gravity Gradient}

Even though the CubeSat is nominally in freefall and in a vacuum, it will be subject to external perturbations that can cause the beam to wander~\cite{Udrea2013}. The relative magnitude of these forces depends on the orbital altitude. In LEO, the main effects will be due to residual atmospheric density, gravity gradient, and magnetic interactions. We may ignore the effect of solar radiation pressure as transmission experiments will be conducted in eclipse. The interaction of any residual magnetic moment of the satellite with the Earth's field will cause a bias torque. The gravity gradient will produce a tidal force leading to a restoring torque aligning the satellite with its long axis in the nadir direction. Both magnetic dipole and gravity gradient effects can be minimized by careful design of the CubeSat. These quasi-static influences are easily compensated by the ADCS system and should have minimal effect on the fine-pointing mechanism.

\subsubsection{Atmospheric Buffeting}

A source of random torque will be the effect of residual atmospheric density in low Earth orbit~\cite{lyle1971spacecraft}. A CubeSat at this altitude experiences free-molecular flow and is potentially subject to buffeting, especially from cross-track winds at high latitudes~\cite{garcia2014atmospheric}. The induced torque due to imbalanced forces can be minimized by locating the centre of gravity close to the centre of pressure when in the relevant orientation to reduce the moment arm. During the quantum transmission phase, the satellite is oriented to present the minimal projected area, i.e. the 2U-3U faces. The lack of deployable solar arrays is advantageous from this respect. Data from the MinXSS mission deployed from the ISS constrains the effect to below $50\mu rad$ for the 3U CubeSat with deployables~\cite{mason2016minxss}.

\subsubsection{Vibration}

The momentum wheels are a potential source of vibration than could affect the pointing accuracy of the beam steering system. A key development goal would be to characterize ADCS hardware bias off-sets and noise spectra to assist in performance modelling~\cite{hegel2016flexcore}, e.g. using coloured noise instead of white noise normally assumed in most simulations and incorporating reaction wheel essential spin-axis instabilities that they may exhibit. TU Delft have experience with these challenges through their CubeSats projects Delfi-C3 and DelfiN3Xt. The BRITE CubeSat missions for photometry also require highly accurate and stable precision pointing systems and have studied the effect of ADCS vibration~\cite{BRITEADCS}. Through careful component selection and modelling, the effect of wheel imbalances can be minimized and in this way TUGSat-1 (BRITE-Austria) has achieved an in-orbit demonstration of $50\mu rad$ using only body pointing and without beam steering~\cite{weiss2014brite,sarda2016three}.

To minimise vibration and enhance spacecraft agility, the ADCS can be operated in a zero momentum mode where the speed of the wheels is low~\cite{steyn1999attitude}. The operational procedure would be dump excess momentum using the MTQs prior to the transmission phase where the wheels are used to provide attitude control. This requires the use of micro-reaction wheels that can support this mode of operation, especially repeated zero-crossings.

\subsubsection{Atmospheric scattering, absorption, and distortion}

The passage of light through the atmosphere is subject to various effects that will reduce the intensity of the received signal. The main sources of error are scattering and absorption of light from the beam and beam wander due to turbulence. Scattering and absorption can be minimized by choice of wavelength and operating conditions. Light at $800nm$ is transmitted through clear air with moderate absorption or scattering~\footnote{For example, $70\%$ of light will be transmitted from space to sea level at $20^\circ$ from zenith~\cite{clark1999spectroscopy,modtran}}. Cloud or other particulates will degrade the channels so clear conditions will be necessary for transmission experiments.

Wavefront distortion due to spatio-temporal variation of refractive index due to turbulence leads to beam wander~\footnote{As the beam is small, we are mainly concerned with wavefront tilt rather than higher order perturbations so more complex adaptive optics is not required.}, the same effect that limits astronomical seeing. The shower curtain effect~\cite{yura1979signal,dror1998experimental} means that the beam wander for an orbit to ground transmission will be smaller that for a ground to space transmission for the same atmospheric turbulence~\cite{bourgoin2013comprehensive}. Since the optical beacon and downlink photons take similar paths, separated by the velocity aberration angle, this will partially cancel out the effect of beam wander as long as the OBT detection and BSM bandwidth is greater than the timescale of the turbulence. The magnitude of the nearly common path rejection will depend on the size of the turbulent cells compared with the beam displacement between up and down-going beams which, at the top of the stratosphere, is a maximum of $3m$ at zenith and reduces to zero as the satellite approaches the horizon.

An additional effect is dispersion of the different wavelengths of the beacon and downlink photons leading to angular differences as they pass through the atmosphere. This will lead to a quasi-static correction to the computed velocity aberration point-ahead of the downlink from the observed OBT position, but also variation in the respective deflections due to turbulence that will be more difficult to compensate. The static dispersion of $\sim 4\mu rad$ displacement between the upgoing $532nm$ and downcoming $800nm$ beams is greatest at low elevations~\cite{Aoki2014Manual}. A correction can be included with the point-ahead compensation~\footnote{The velocity aberration is maximal when the dispersion displacement is minimum at zenith, and vice versa near the horizon}.

\section{Missions}

CQuCoM calls for two missions, the first to derisk the pointing mechanism with a high brightness transmission source that can also be used for WCP QKD, and a second mission to distribute entanglement between space and ground. The mission profiles for both are broadly similar. A launch broker such as Nanoracks~\cite{nanoracks} will be contracted to handle orbital deployment~\footnote{Spaceflight Industries~\cite{spaceflight} can broker deployment on a variety of launchers and in different orbits allowing for some flexibility on mission planning should the ISS orbit not be suitable.}. The CQuCoM satellites will first be carried up to the ISS on a regular resupply mission (Dragon, Cygnus, HTV, ATV, Progress and Soyuz) and then deployed into orbit using the NanoRacks CubeSat Deployer (NRCSD) mounted upon the Japanese Experimental Module Remote Manipulator System (JEMRMS).

\subsection{In-Orbit Operations}

After switch-on and detumbling, the satellite will initiate basic housekeeping procedures such as charging the batteries, establishing contact with ground control, and monitoring onboard systems. The performance of the ADCS will be verified and tests of ground target tracking can be performed in daylight using the OBT imager with the narrow bandpass filter removed. An option for an adjustable defocus for the OBT will be investigated for imaging purposes as opposed for centroiding.

Initial night passes over the OGS will verify both satellite and beacon acquisition and tracking as well as the operation of the realtime telemetry downlink. The first mission with a tunable WCP source will allow sighting-in of the OBT/BSM, in particular to check that alignment of the incoming and outgoing beam paths have not deviated from that determined by pre-flight ground tests, e.g. by using a spiral search pattern of the BSM. For the second mission with the entangled source, it may still be possible to pick out the single photon flux from the satellite during a slow spiral pattern assuming small shifts in the boresight alignment. The results of the first mission will be vital in determining the effect of launch and orbital environmental conditions on the alignment.

Once the in-orbit optical system parameters have been calibrated, quantum transmission tests can begin. These will be conducted in eclipse (local night) when weather conditions are clear and the orbital track passes close enough to the OGS, rising at least $30^\circ$ above the horizon. As the satellite begins to rise above the horizon, it will use ADCS to point the telescope towards the expected position of the OGS. Conversely, the OGS will track the satellite as it appears. Laser rangefinding pulses will provide precise position information for the OGS and it can begin transmitting the laser beacon. The satellite uses the beacon to operate the fine-pointing system. Once the OBT is locked onto the beacon, the source can start transmitting quantum signals to the OGS. Telemetry from the satellite to the OGS will transmit orientation information from the ADCS system allowing the alignment of the OGS polarization measurement bases with those being transmitted. The entanglement source has the option of actively adjusting its own polariser analysis settings based on onboard orientation information leaving the OGS settings fixed.

Synchronisation of CubeSat source and OGS receiver events can be performed via GPS timing signals and post-transmission processing using the ranging information determined from the retroreflected laser pulses. Synchronisation can also be performed through modulation of the beacon signal and a separate photodiode. To reduce the amount of information needed to be stored and transmitted by the CubeSat, the OGS can communicate detection events, only the corresponding onboard data (WCP random signal settings or detection events for the entangled source) in the temporal vicinity need can be retained. The OGS detection rate (signal plus background and dark counts) will be in the range $10^3 s^{-1}$ to $10^4 s^{-1}$ due to channel losses and, in principle, only the coincident events need be processed or downloaded.

If the notification of OGS events is done in realtime to the CubeSat, either through the S-band uplink or laser beacon, this minimizes the total amount of onboard storage than needs to be provided~\footnote{E.g. a ring buffer could be used to temporarily store onboard signals or timing data and only coincident events would be copied out to main storage. In practice, a range of data around the OGS events would be copied out to guard against timing inaccuracies and to assist post-transmission analysis and synchronization.}. However, for scientific purposes it would be beneficial to store the entire onboard record during a pass and download for ground analysis. Due to the high source rate, this will result in several $GB$ of data that needs to be downlinked~\footnote{For a $400s$ quantum transmission pass (which is optimistic), a $100MHz$ WCP source will require $\sim 10^{11}$ bits ($4\times 10^{10}$ signals and $4 bits/signal$ if using decoy states). For a $5Mpcs$ continuously pumped entangled photon source, we require $20 bits$ timing information per detection event leading to $4\times 10^{10}$ bits per pass. We thus assume $20GB$ of onboard signal or timestamp data per pass.}. High speed X-band CubeSat transmitters are now commercially available and in use allowing large amounts of data to be downloaded from orbit. The company Planet reports $4.2GB$ downloaded during a typical groundstation pass from 3U CubeSats using COTS communications equipment~\cite{klofas2016}. With 3 groundstations and several passes per station per day, the data generated from a single quantum transmission experiment should downloadable within a day. The S-band uplink can be used for post-quantum-transmission reconciliation and processing of the coincident event data, e.g. sifting, error correction, and privacy amplification, if required for QKD demonstration.

\subsection{Decommissioning}
\label{sec:deorbit}

\begin{figure}
\includegraphics[width=0.95\textwidth]{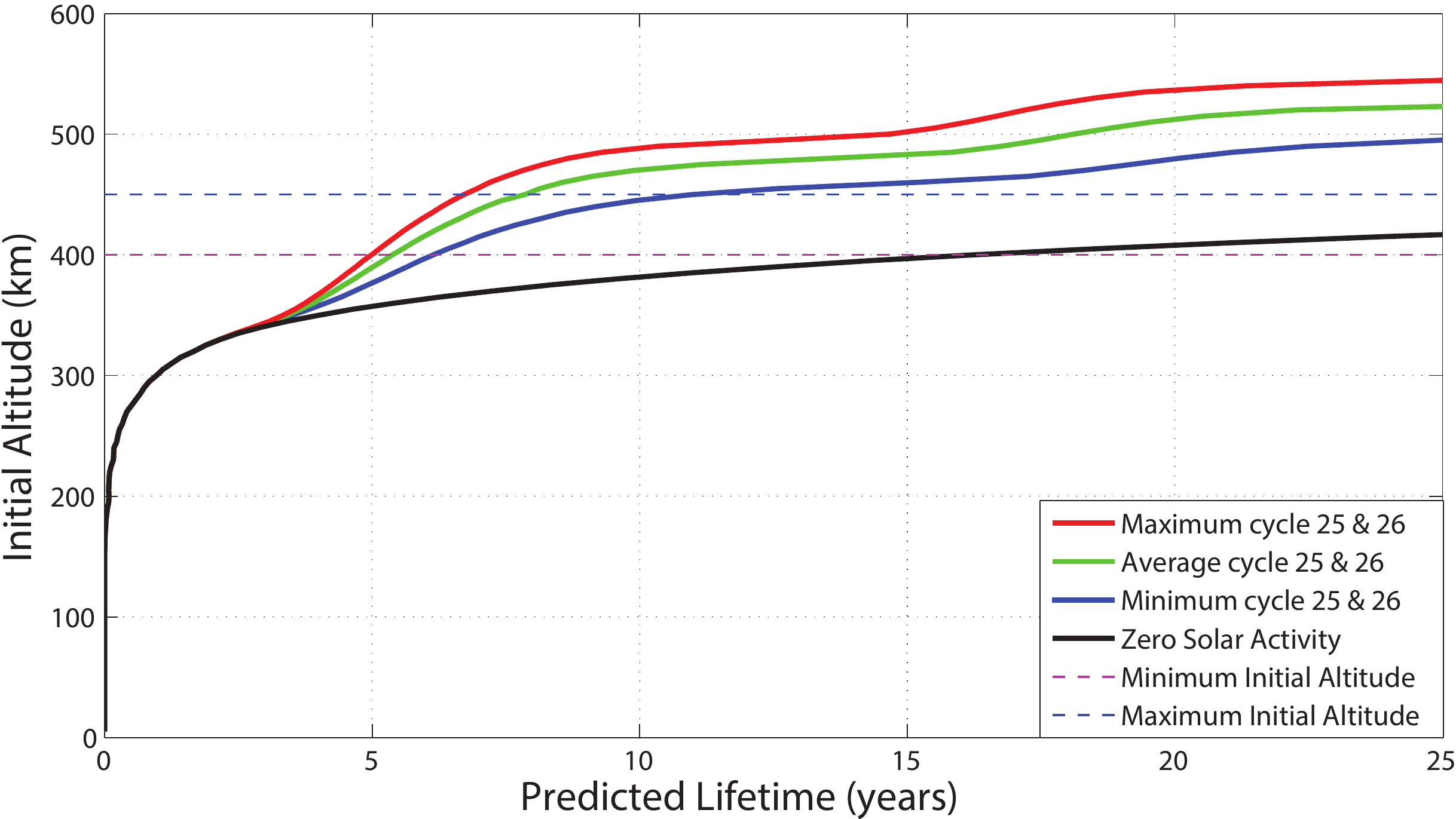}
\caption{Deorbit analysis for a $10kg$ 6U CubeSat in minimum drag configuration. As atmospheric drag depends strongly on solar activity and consequent atmospheric expansion, we calculate orbital lifetimes for consecutive minimum (very low intensity, solar max of 140SFU), average (moderate intensity, solar max of 190SFU) or maximum (high intensity, solar max of 230SFU) solar cycles following the current solar cycle. An extended period of no solar activity, much like the Maunder Minimum event, is included for completeness. The representative solar cycles used herein were derived from historical solar cycle data. The ISS orbit varies and is periodically reboosted with orbit raising manoeuvres to combat orbital decay, the range of altitudes is indicated by the dashed horizontal lines and is derived from the standard operating altitudes of the ISS and are thus subject to change.
}
\label{fig:deorbit}
\end{figure}


Space debris is a major issue for any satellite mission and satellites should be designed to de-orbit within 25 years of launch~\cite{deorbit} and by design CQuCoM should meet this directive. If an orbital altitude beyond $500km$ is chosen, either due to launch opportunity or reduction in atmospheric buffeting, then meeting the 25 year de-orbit directive may require additional mechanisms, increasing mass, developmental effort, and cost. A deployment below $500km$ simplifies the decommissioning task as the satellite will passively de-orbit in a relatively short period. A typical CubeSat deployed at the altitude of the International Space Station will have an orbital lifetime ranging from months to a few years.

In order to demonstrate the potential de-orbit period of a $10kg$ 6U CubeSat, it is assumed that the CubeSat would be in minimum drag configuration (i.e. minimum projected area) and that the CubeSat would be launched from the ISS in Q1 of 2018. The method developed by Kerr and Macdonald~\cite{kerr2015general} was used to calculate the de-orbit period and the results are presented in Fig.~\ref{fig:deorbit}. It can be seen that if the CubeSat were deployed at the maximum ISS altitude of $450km$, even in the case where cycles 25 and 26 are of very low intensity, the de-orbit period is approximately 11 years. However, in the case of an extended period of zero activity and deployment from $450km$, the CubeSat lifetime will exceed the 25 year best practice rule. In periods of low solar activity the ISS can maintain a lower altitude but in periods of high solar activity, a higher altitude is chosen to reduce drag. However an upper limitation on the orbit exists due to the operating limits of the spacecraft which rendezvous with the ISS. In practice, we would expect that during periods of low or no solar activity the ISS would be at the lower range of its altitude range and the 25 year de-orbit limit can be met.

\section{Conclusion and Outlook}

CubeSats offer the potential to accelerate the development of quantum technologies in space by offering reliable, and cost-effective platforms for conducting in-orbit technology demonstrations. The cost-effectiveness of CubeSats is derived from the standard containers used to ship and deploy CubeSats. This has led to the ability to share launch costs between a large number of users. At the same time, advances in micro-electronics and RF communication have enabled many advanced experiments to be operable remotely, using only COTS components. Together, these advances have made in-orbit experiments accessible to university groups and consortia that were not space users, even a decade previously.

Some physical parameters, such as aperture-size and diffraction-losses, that are associated with optical systems are expected to become relatively more important requirement drivers of an experiment system design. However, this is an advantage as it means that from a systems engineering perspective, there is now greater flexibility in how to put together a space-based quantum experiment. With these positive developments, we can look forward to more nanosatellite sized experiments that act either as path-finders for more advanced experiments, or to actually execute the actual scientific experiments. The CQuCoM proposal combines the aforementioned advantages for advanced missions that are at the leading edge of small satellite capabilities.


\section*{CQuCoM Consortium}

The CQuCoM consortium consists of:
\begin{description}
\item[University of Strathclyde] Co-ordination, Mission Operations
\item[Austrian Academy of Sciences] Mission Planning, Scientific Oversight
\item[Clyde Space Ltd] Platform Engineering and Testing
\item[Technical University of Delft] Optical Design, ADCS Design and Algorithms
\item[Ludwig-Maximilian University] Fine-pointing system and WCP Source
\item[University of Padua] Optical Ground Station (MLRO), in collaboration with ASI - Italian Space Agency
\item[National University of Singapore] Entanglement Source and Data Handling
\end{description}


\begin{backmatter}

\section*{Competing interests}
  The authors declare that they have no competing interests.

\section*{Author's contributions}
The original CQuCoM concept was conceived jointly by DO, AL, and SG. The other authors are involved either in the development of the mission concept, subsystems, or procedures. All authors read and approved the final manuscript.

\section*{Acknowledgements}
DKLO acknowledges QUISCO (Quantum Information Scotland Network). DKLO and AL received funding from the European Union’s Seventh Framework Programme for research, technological development and demonstration under grant agreement No.611014 (CONNECT2SEA-4). AL acknowledges support from the National Research Foundation Singapore (NRF-CRP12-2013-02). EC acknowledges support from the Swiss National Science Foundation. 

\bibliographystyle{bmc-mathphys} 

\end{backmatter}
\end{document}